\documentclass[prb,preprint,showpacs,amsmath,amssymb,superscriptaddress]{revtex4}
\usepackage{epsfig}
\usepackage{graphicx}
\usepackage{eso-pic,everyshi,ifthen,calc}
\usepackage{amsmath,amssymb}
\usepackage{pdfpages}

\begin{document}

\title{Non-thermal laser induced precession of magnetization in ferromagnetic semiconductor (Ga,Mn)As}

\author{P.~N\v{e}mec}
\author{E.~Rozkotov\'a}
\author{N.~Tesa\v{r}ov\'a}
\author{F.~Troj\'anek}
\affiliation{Charles University in Prague, Faculty of Mathematics and Physics, Ke Karlovu 3, 121 16 Prague 2, Czech Republic}

\author{K.~Olejn{\'{i}}k}
\affiliation{Institute of Physics ASCR, v.v.i., Cukrovarnick\'a 10, 162 53 Praha 6, Czech Republic}

\author{J.~Zemen}
\affiliation{Institute of Physics ASCR, v.v.i., Cukrovarnick\'a 10, 162 53 Praha 6, Czech Republic}

\author{V.~Nov{\'ak}}
\affiliation{Institute of Physics ASCR, v.v.i., Cukrovarnick\'a 10, 162 53 Praha 6, Czech Republic}

\author{M.~Cukr}
\affiliation{Institute of Physics ASCR, v.v.i., Cukrovarnick\'a 10, 162 53 Praha 6, Czech Republic}

\author{P.~Mal\'y}
\affiliation{Charles University in Prague, Faculty of Mathematics and Physics, Ke Karlovu 3, 121 16 Prague 2, Czech Republic}

\author{T.~Jungwirth}
\affiliation{Institute of Physics ASCR, v.v.i., Cukrovarnick\'a 10, 162 53 Praha 6, Czech Republic}
\affiliation{School of Physics and Astronomy, University of Nottingham, Nottingham NG7 2RD, United Kingdom}

\date{\today}
\pacs{75.50.Pp, 76.50.+g, 78.20.Ls, 78.47.-p}

\maketitle

{\bf
Non-thermal laser induced spin excitations, recently discovered in conventional oxide and metal ferromagnets,\cite{Kimel:2005_b,Stanciu:2007_a,Bigot:2009_a} open unprecedented opportunities for research and applications of ultrafast optical manipulation of magnetic systems.  Ferromagnetic semiconductors, and (Ga,Mn)As in particular, should represent  ideal systems for exploring this new field.  Remarkably, the presence of non-thermal effects has remained one of the outstanding unresolved problems in the research of ferromagnetic semiconductors to date.\cite{Oiwa:2005_a,Qi:2007_a,Qi:2009_a,Rozkotova:2008_a,Rozkotova:2008_b,Hashimoto:2008_a,Hashimoto:2008_b,Kobayashi:2010} Here we demonstrate that coherent magnetization dynamics can be excited in (Ga,Mn)As non-thermally by a transfer of angular momentum from circularly polarized femtosecond laser pulses and by a combination of non-thermal and thermal effects  due to a transfer of energy from laser pulses. The thermal effects can be completely suppressed in piezo-electrically controlled samples.  Our work is based on pump-and-probe measurements in a large set of (Ga,Mn)As epilayers and on systematic analysis of circular and linear magneto-optical coefficients. We provide microscopic theoretical interpretation of the experimental results.}

(Ga,Mn)As and related ferromagnetic semiconductors are potentially ideal testbed materials for exploring laser induced excitations of magnetization. Their direct-gap band structure allows for strong optical excitations of the electronic system, the photo-carriers can directly interact with magnetic moments via strong exchange coupling, the carrier mediated ferromagnetism produces large and tunable magnetic and magneto-optical effects, and the relatively simple band structure is favorable for identifying  microscopic physical origins of the phenomena. Femtosecond laser pulse induced precession of magnetization in ferromagnetic (Ga,Mn)As has been recently reported by several groups. \cite{Oiwa:2005_a,Qi:2007_a,Qi:2009_a,Rozkotova:2008_a,Rozkotova:2008_b,Hashimoto:2008_a,Hashimoto:2008_b}  Since  no dependence on the helicity of the pump laser beam has been identified, the precession in these experiments is a consequence of an impulsive change of the magnetic anisotropy in (Ga,Mn)As. To discern the presence of non-thermal effects under these circumstances is not straightforward. The magnetic anisotropy can in principle change not only due to the photo-carriers but also due to the transient increase of temperature.

In Ref.~\onlinecite{Qi:2009_a}, the magnetization precession was induced by laser pulses of a relatively weak intensity $\sim 1-10\mu$Jcm$^{-2}$ and ascribed to heating effects. On the other hand, in Refs.~\onlinecite{Oiwa:2005_a} and \onlinecite{Hashimoto:2008_b} the observed precession triggered by comparably weak laser intensities was attributed to the direct effect of photo-injected holes on the magnetic anisotropy. Apart from these competing views,  the photo-carrier based interpretation  cannot be  reconciled with the theoretical and experimental understanding of static magneto-crystalline anisotropies in (Ga,Mn)As.\cite{Hashimoto:2008_b} The interpretation  would imply that measured changes of the orientation of the easy-axis in materials with equilibrium hole densities $\sim10^{20}-10^{21}$~cm$^{-3}$ are induced by photo-injected holes of density as low as  $\sim10^{16}$~cm$^{-3}$. To compare with, e.g., electrical gating experiments,\cite{Chiba:2008_a,Owen:2008_a} we recall that changes in the magnetic anisotropy are detected for field-induced hole accumulation or depletion of at least $\sim10^{18}-10^{19}$~cm$^{-3}$. We also point out that the original interpretation in Ref.~\onlinecite{Oiwa:2005_a} in terms of photo-carrier induced changes  of in-plane anisotropy fields was subsequently revised by assuming\cite{Hashimoto:2008_b,Kobayashi:2010} out-of-plane tilts of the easy-axis. Since out-of-plane easy-axis rotation is not consistent with measured trends in static magnetic anisotropies of the considered (Ga,Mn)As/GaAs materials this again illustrates that the non-thermal laser induced magnetization precession due to optically generated carriers in ferromagnetic semiconductors has remained an attractive yet unproven concept. In this paper we directly observe non-thermal excitation of the magnetization in (Ga,Mn)As by detecting the dependence of the magnetization dynamics on circular polarization of the pump laser pulse, i.e., on the spin-polarization of excited photo-carriers that is directly connected with a transfer of angular momentum from circularly-polarized laser pulses. We also observe a clear evidence of the non-thermal effect in the component of the dynamics which is independent of the helicity of the pump beam. We find that photo-carriers can dominate  this magnetic anisotropy driven mechanism at high pump laser intensities.

For our study we utilize a set of (Ga,Mn)As/GaAs materials with individually optimized molecular-beam-epitaxy  and post-growth annealing procedures  for each nominal Mn doping in order  to  minimize the density of compensating defects and other unintentional impurities and to achieve high uniformity of the epilayers. Nominal Mn-dopings in this set of 20~nm thick ferromagnetic (Ga,Mn)As epilayers span the whole range up to $\sim 14$\% (corresponding to $\sim 8$\% of uncompensated Mn$_{\rm Ga}$ impurities) with ferromagnetic transition temperatures reaching 190~K.  All samples within the series have  reproducible characteristics with the overall trend of  increasing Curie temperature, increasing hole concentration, and increasing magnetic moment density with increasing nominal Mn doping. The magnetic anisotropy characteristics systematically vary with doping and further {\em in situ} electrical control is achieved by an attached PZT piezo-stressor that generates a tensile (compressive) strain if a positive (negative) piezo-voltage is applied.  Detail descriptions of our (Ga,Mn)As epilayers and of the preparation and characterization of the PZT/(Ga,Mn)As hybrid structures are given in Ref.~\onlinecite{Jungwirth:2010_b,Rushforth:2008_a,Ranieri:2008_a} and in the Supplementary Information.

Laser-pulse induced dynamics of magnetization is investigated by the pump-and-probe magneto-optical technique.\cite{Wang:2006_d,Kirilyuk:2010_a} A schematic diagram of our  experimental set-up is shown in Fig.~1(a). The output of a femtosecond laser is divided into a strong pump pulse and a weak probe pulse that are focused to the same spot on the investigated sample. The impact of the pump pulse, which is spectrally tuned above the band gap of (Ga,Mn)As, modifies the properties of the sample. The resulting changes of the magneto-optical response of the sample, i.e. the pump-induced change of the probe rotation or ellipticity, are measured by the time-delayed probe pulse. Our pump-pulse intensities are in the range of 4 -- 250~$\mu$Jcm$^{-2}$ corresponding to injected photo-carrier densities of 2.4 -- 150$\times 10^{17}$~cm$^{-3}$.

\begin{figure}[h!]
\vspace*{-2cm}
\includegraphics[width=.78\columnwidth,angle=0]{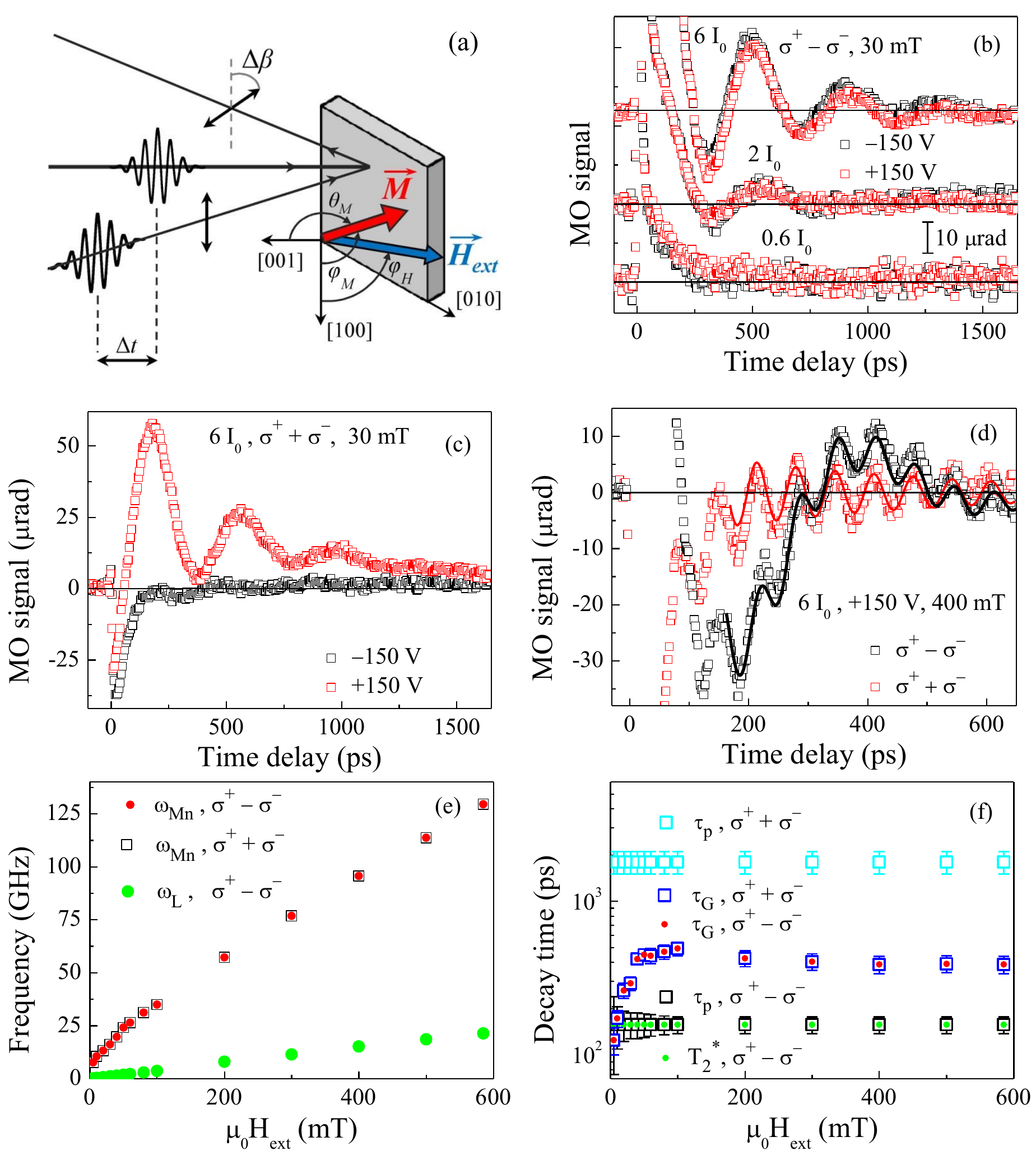}
\vspace*{-0.5cm}
\caption{{\protect\footnotesize Laser pulse-induced precession of magnetization measured at 35~K in a ferromagnetic (Ga,Mn)As epilayer with nominal 3.8\% Mn-doping  attached to a piezo-stressor. (a) Schematic diagram of the experimental set-up. The orientation of magnetization in the sample is described by the polar $\varphi_M$ and azimuthal $\theta_M$ angles, external magnetic field $H_{ext}$ is applied in the sample plane at an angle $\varphi_H$. The rotation of the polarization plane $\Delta\beta$ of the reflected linearly polarized probe pulse is measured as a function of the time delay $\Delta t$ between pump and probe pulses with the photon energy 1.63~eV. (b) -- (d) Dynamics of the magneto-optical signal measured for the helicity-dependent ($\sigma^+-\sigma^-$) and helicity-independent ($\sigma^++\sigma^-$) excitations. Pump laser intensities (with $I_0=7$~$\mu$Jcm$^{-2}$), piezo-voltages, and magnitudes of external field applied at an angle $\varphi_H=115^\circ$ are specified in each panel. Input polarization orientation of the probe pulses is $\beta= 25^\circ$ (measured from the [100] crystal direction). (e) Frequency of the precessing ferromagnetic Mn moments ($\omega_{Mn}$) and Larmor precession frequency of electrons ($\omega_L$) and (f) pulse-function ($\tau_p$), Gilbert ($\tau_G$), and electron spin ($T^\ast_2$) damping times as a function of $H_{ext}$ for $\varphi_H=115^\circ$, piezo-volatge +150~V, an excitation intensity $6I_0$. }
}
\label{fig1}
\end{figure}

A direct evidence of the non-thermal laser induced spin-precession in ferromagnetic semiconductor (Ga,Mn)As is shown in Fig.~1(b) where we demonstrate the dependence of the magnetization dynamics on the helicity of circular polarization of the pump laser beam. The measurements are performed on a (Ga,Mn)As epilayer with nominal 3.8\% Mn-doping (Curie temperature $T_c=96$~K). The magnitude of the oscillatory signal is proportional to the intensity of the pump laser pulse, i.e., to the density of injected spin-polarized photo-carriers. In Fig.~1(c) we plot the component of the dynamical signal which is independent of the helicity of the circular polarization of the pump laser beam. It corresponds to the magnetic anisotropy driven mechanism, namely to the laser induced tilt of the magnetic easy-axis due to transient heating and photo-excitation of unpolarized carriers. 

The distinct nature of  mechanisms responsible for the dynamics observed in Figs.~1(b) and (c)  is highlighted by measurements at different electric fields applied to the attached piezo-stressor. In the case of the helicity-dependent laser excitation, the stressor is expected to have a negligible effect on the spin density of excited photo-carriers and, therefore, also on the induced magnetization dynamics. On the other hand, the applied stress can strongly modify magnetic anisotropies\cite{Rushforth:2008_a,Ranieri:2008_a,Zemen:2009_a} which drive the helecity-independent dynamics. Consistent with these expectations there is no significant dependence of the laser induced dynamics on the piezo-voltage seen in Fig.~1(b) and a strong dependence seen in Fig.~1(c). Remarkably, we succeeded to completely suppress the helicity-independent signal at large negative piezo-voltages and external magnetic fields in the range of 10 -- 40~mT applied at an angle $\varphi_H=115^\circ$ from the [100]-axis, as shown in Fig.~1(c). At these external field conditions the observed laser induced dynamics of magnetization is entirely due to the non-thermal effect of spin-polarized photo-carriers.

We fit the measured pump-induced dynamical change of the magneto-optical signal, $\delta MO$, by the equation,
\begin{equation}
\delta MO(t)=A\cos(\omega_{Mn} t+\Delta)e^{-t/\tau_G}+Ce^{-t/\tau_p}+D\cos(\omega_Lt)e^{-t/T^\ast_2}\;
\label{mo_dyn}
\end{equation}
where $A$ and $C$ are the amplitudes of the oscillatory and pulse function, respectively, $\omega_{Mn}$ is the frequency of precessing ferromagnetic Mn-moments, $\Delta$ is a phase factor, $\tau_G$ is the Gilbert damping time, and $\tau_p$ is the pulse function decay time.\cite{Rozkotova:2008_b} Apart from the precessing ferromagnetic Mn moments seen in both helicity-dependent and helicity-independent signals, the helicity-dependent signal has another lower frequency component due to the photo-excited spin-polarized conduction band electrons (see Fig.~1(d)). This signal is described by the last term in Eq.~(\ref{mo_dyn}) where $D$, $\omega_L$, and $T^\ast_2$ are the amplitude, Larmor precession frequency, and transverse spin coherence time of electrons, respectively. 

In Fig.~1(e) we show that the helicity-dependent and helicity-independent oscillating signals have identical frequencies $\omega_{Mn}$. They correspond to the Landau-Lifshitz-Gilbert (LLG) dynamics of ferromagnetic Mn moments with the g-factor $g_{Mn}=2$ in equilibrium magnetic anisotropy fields which were independently determined from static magnetization measurements. The Larmor frequency  $\omega_L$ corresponds to the g-factor $|g_e|=0.4$ of the conduction band electrons. Note that the precession of photo-generated holes in the valence band is not observed due to the short, sub-picosecond spin lifetime of the strongly spin-orbit coupled valence band holes. 

The exchange coupling between conduction band s-electron spins and the Mn d-electron local moments 
is $J_{sd}\sim10$~meV~nm$^3$. Assuming the electron spin density $s$ exited by circularly polarized light in a  clean GaAs semiconductor as the upper bound for the corresponding photo-electron spin density in (Ga,Mn)As, the expected fields acting on the ferromagnetic Mn moments, $J_{sd}s/g_{Mn}\mu_B$, can be as high as $\sim20$~mT per 10$^{18}$cm$^{-3}$ electron density. These fields can readily account for the observed laser induced precession of the ferromagnetic Mn moments. Consistent with this interpretation, the characteristic decay time $T^\ast_2$ of the photo-electron spin polarization observed in the helicity-dependent signal coincides with the decay time $\tau_p$ of the pulse function in the Mn moment dynamical signal, as shown in Fig.~1(f). The pulse function corresponds to the time-dependent tilt of the vector around which the ferromagnetic moments precess. The straightforward mechanism of the light absorption generating spin-polarized carriers which directly act on the magnetic moments via the exchange field has not been reported prior to our work and illustrates the unique potential of ferromagnetic semiconductors in the field of non-thermal laser induced magnetization dynamics. In the helicity-independent signal the significantly larger decay time of the pulse function  (see Fig.~1(f)) corresponds to the return of the laser induced tilt of the magnetic easy-axis to the equilibrium position due to the decay of the transient heating and photo-excitation of unpolarized carriers. 

In Fig.~2 we present a detailed analysis of the pulse function and of the oscillating part of the magneto-optical signal. Magnetization dynamics data measured using different orientations of the linear polarization of the probe pulses for the helicity-dependent and helicity-independent excitations are shown in Figs. 2(a) and (b), respectively. The magneto-optical signal represents the rotation of the polarization plane $\Delta\beta$ of the reflected linearly polarized probe beam. It comprises the signal due to the out-of-plane motion of the magnetization, which is sensed by the polar Kerr effect (PKE), and the signal due to the in-plane component of the ferromagnetic moment, which is sensed by the magnetic linear dichroism (MLD).\cite{Kimel:2005_a} These two contributions can be experimentally separated by their polarization dependence; PKE does not depend on the probe input polarization angle $\beta$ while MLD is a harmonic function of $\beta$. The static magneto-optical signal can be written in a form (for more details see Supplementary Information), 
\begin{equation}
MO^{stat}=P^{PKE}\cos\theta_M+P^{MLD}\sin\theta_M\sin2(\varphi_M-\beta)\;.
\label{mo_stat}
\end{equation}
Here $\theta_M$ and $\varphi_M$ are the polar and azimuthal angles of the equilibrium magnetization and  $P^{PKE}$ and $P^{MLD}$ are the PKE and MLD coefficients determined for the particular (Ga,Mn)As material from static magneto-optical measurements. For small excitations from equilibrium, the amplitude of the pulse function is a sum of the PKE and MLD contributions, $C=C^{PKE}+C^{MLD}$, and reflects a change of the ferromagnetic moment orientation (shift signal) and a change of the magnitude of the moment (demagnetization signal):
\begin{eqnarray}
C&=&C_{shift}+C_{demag} \nonumber \\
C_{shift}&=& -\delta\theta P^{PKE}+\delta\varphi P^{MLD}2\cos2(\varphi_M-\beta) \nonumber \\
C_{demag}&=&-\delta pP^{MLD}\sin2(\varphi_M-\beta)\;,
\end{eqnarray}
where $\delta\theta$ and $\delta\varphi$ denote changes of the polar and azimuthal angles of the magnetization due to the pump pulse and $\delta p$ is the demagnetization factor. Analogous expressions can be derived for the amplitude of the oscillatory part of the dynamic magneto-optical signal, $A^2=(A^{MLD})^2+(A^{PKE})^2$ (for more details see Supplementary Information). 

\begin{figure}[h!]
\vspace*{0cm}
\includegraphics[width=0.9\columnwidth,angle=0]{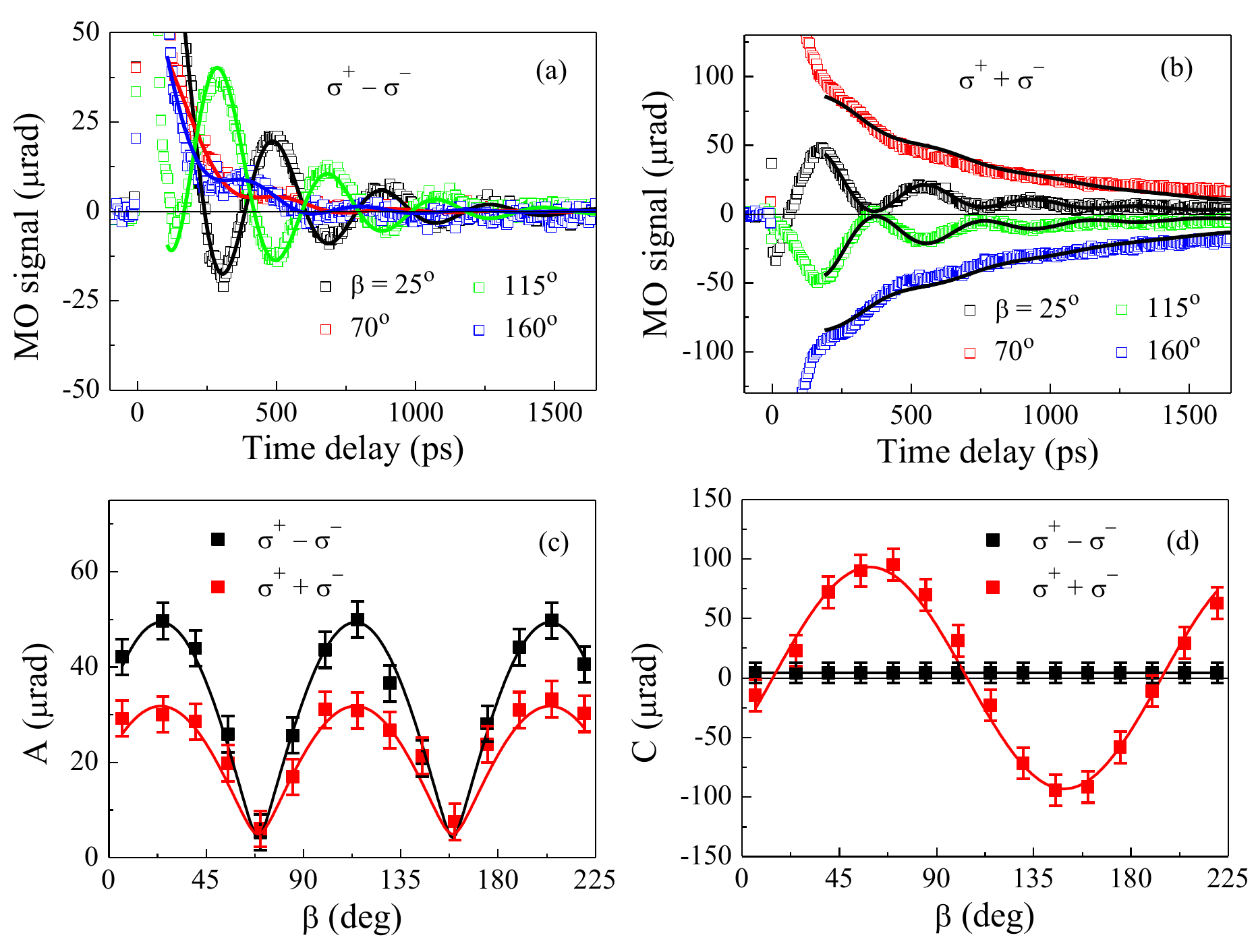}
\caption{{\protect\footnotesize (a),(b) Dynamics of the magneto-optical signal measured by probe pulses with different $\beta$ for the helicity-dependent ($\sigma^+-\sigma^-$) and helicity-independent ($\sigma^++\sigma^-$) excitations, in the 3.8\% Mn-doped (Ga,Mn)As, at temperature  35~K, piezo-voltage +150~V, excitation intensity $6I_0$, $\mu_0H_{ext}=30$~mT, $\varphi_H=115^\circ$. (c),(d) Polarization dependence of the amplitudes of the oscillatory part ($A$) and of the pulse function ($C$). Lines are fits described in main text and Supplementary Information.}}
\label{Fig2}
\end{figure}
From the measured data we immediately see that the dependence of the pulse function on $\beta$ is very weak for the helicity-dependent signal (Fig.~2(a)) while it is much stronger for the helicity-independent signal (Fig. 2(b)). This shows that the non-equilibrium vector along which the magnetization precesses is tilted in the out-of-plane direction for the helicity dependent excitation and in the in-plane direction for the helicity-independent signal. Since we excite with the normal incidence laser beam, the former observation is another confirmation that photo-carriers with their spins polarized in the out-of-plane direction act on the magnetization in case of the helicity-dependent signal. The latter observation is also reconfirming since our (Ga,Mn)As materials are strong in-plane magnets so the easy-axis rotation induced by the absorption of the laser pulse (i.e. by the transient heating and by the photo-induced change of the hole concentration) can only occur within the epilayer plane. 

We extracted the amplitudes of the  oscillatory part and of the pulse function by fitting Eq.~(\ref{mo_dyn}) to the measured magneto-optical dynamical curves. The resulting $\beta$-dependent $A$ and $C$ for the helicity-dependent and helicity-independent excitations are plotted  in Figs.~2(c) and (d), respectively. $A(\beta)$ has the same oscillatory character for the helicity-dependent and helicity-independent excitations because, in both cases, it corresponds to precessing moments with time-dependent in-plane and out-of-plane components. The constant $C(\beta)$ for helicity-dependent excitation confirms the purely out-of-plane tilt of the non-equilibrium vector along which the magnetization precesses; the oscillatory $C(\beta)$ for the helicity-independent excitation confirms  the in-plane tilt. By fitting $A(\beta)$ and $C(\beta)$ to the derived expressions for the polarization dependences of $A$ and $C$ (see Supplementary Information) we can also obtain the equilibrium easy-axis angle $\varphi_M$, as well as the magnitude and sign of the laser induced tilts $\delta\varphi$ and $\delta\theta$. We will use these extracted values in the discussion below.
 
Having established the presence of non-thermally excited magnetization precession due to circularly polarized pump beam we now proceed to the detail analysis of the helicity-independent signal, i.e., of the signal that is connected with the transfer of energy from laser pulses. We show that the thermal effects saturate at sufficiently large pump pulse intensities, however, the easy axis can be rotated further with increasing laser intensity due to the non-thermal effect of unpolarized photo-holes on the magnetic anisotropy. We have explored the whole series of our (Ga,Mn)As epilayers with nominal Mn-doping ranging from 1.5\% to 14\%.  All materials are in-plane magnets in which the biaxial anisotropy, reflecting the cubic symmetry of the host crystal, competes with an additional uniaxial anisotropy whose  magnitude can be modeled by assuming a uniaxial shear strain.\cite{Zemen:2009_a} As illustrated in Fig.~3(a),  the biaxial anisotropy dominates at very low dopings and the easy axis aligns with the main crystal axis [100] or [010]. At intermediate dopings, the uniaxial anisotropy is still weaker but comparable in magnitude to the biaxial anisotropy. In these samples the two equilibrium easy-axes are tilted towards the [1$\bar1$0] direction and are sensitive to small changes in temperature or hole density. At very  high dopings, the uniaxial anisotropy dominates and the system has one strong easy-axis along the [1$\bar1$0] in-plane diagonal. In agreement with these doping trends in static magnetic anisotropies, we did not observe laser induced precession in the very low and very high doped samples and consistently we observed precessions in samples with intermediate doping. This is illustarted in Fig.~3(b). Observed precession frequencies in the studied set of samples correspond to magneto-crystalline anisotropy fields which are fully consistent with the respective anisotropy fields obtained from magnetization measurements by the superconducting quantum interference device (SQUID). 

\begin{figure}[h!]
\includegraphics[width=1.0\columnwidth,angle=0]{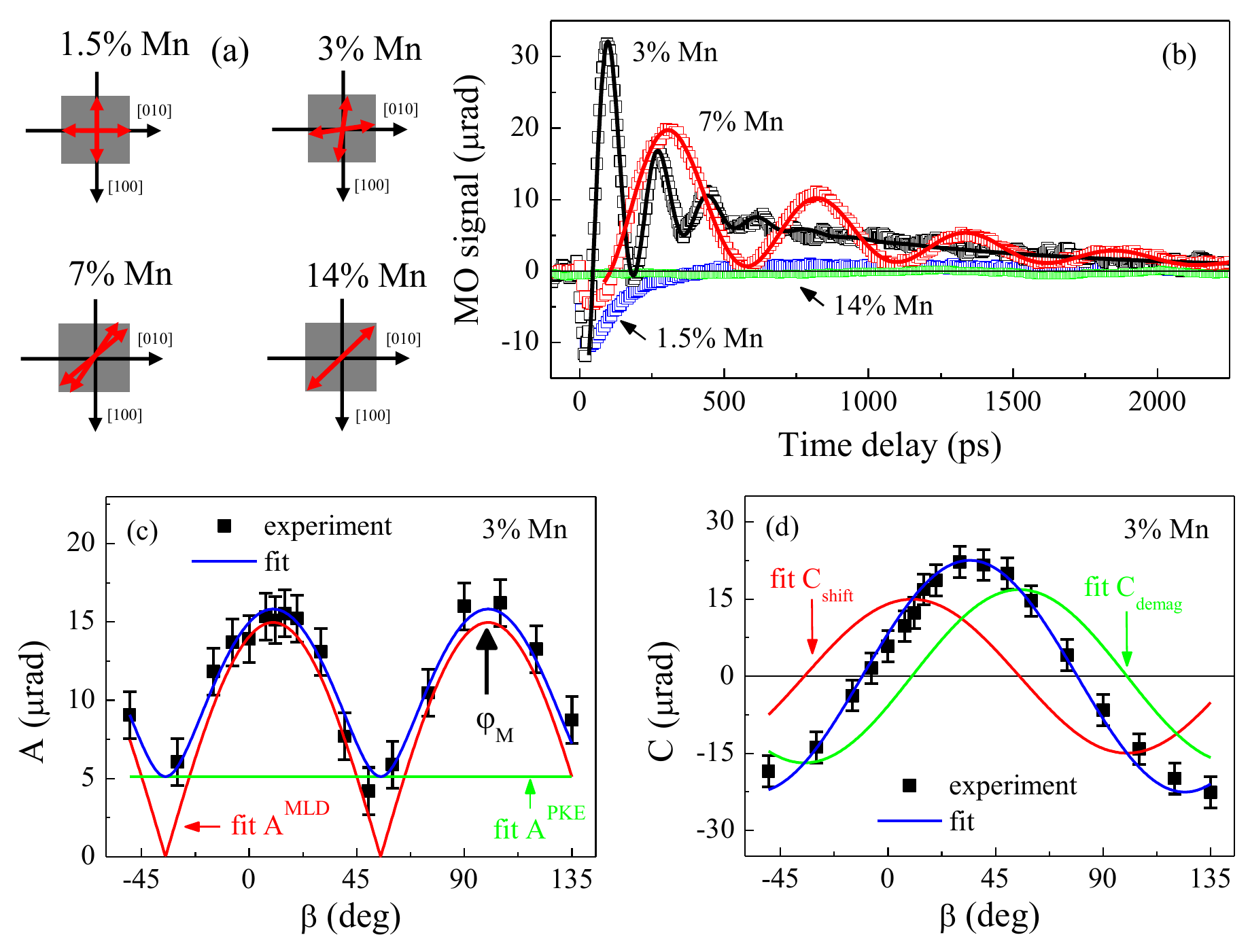}
\caption{{\protect\footnotesize (a)  Schematic diagrams of magnetic easy-axes in (Ga,Mn)As epilayers with different Mn-doping. (b) Dynamics of the magneto-optical signal in (Ga,Mn)As epilayers with depicted Mn-doping measured for the helicity-independent ($\sigma^++\sigma^-$) excitation at temperature 15~K, excitation intensity $I_0$, $H_{ext}\approx 0$, and $\beta=0^\circ$; lines are fits by Eq.~(1). (c),(d) Polarization dependence of the amplitudes of the oscillatory part ($A$) and of the pulse function ($C$) for the 3\% Mn-doped epilayer. Lines and easy-axis angle ($\varphi_M$) are obtained from fits described in main text and Supplementary Information.}
}
\label{Fig3}
\end{figure}
Since our primary interest is in non-thermal effects we have singled out a material from the lower doping end of the set of samples showing laser induced precession of magnetization. This 3\% Mn doped epilayer is still a relatively low hole-density material but with already competing biaxial and uniaxial anisotropies for which we can expect sizable changes due to non-thermal excitation effects at photo-hole densities $10^{18}-10^{19}$~cm$^{-3}$. In Figs.~3(c) and (d) we show an example of the amplitudes of the oscillatory and pulse functions  obtained from the $\beta$-dependent magnetization dynamics measurements at base temperature 15~K and pump laser intensity $I_0=7$~$\mu$Jcm$^{-2}$. From the analysis of $A(\beta)$ and $C(\beta)$ measured at different laser intensities we obtained the  dependence of the laser induced tilt of the easy-axis. Results of this experimental study together with the experimental calibration of the transient temperature change versus laser intensity are summarized in Figs.~4(a)-(c). 

First we plot in Fig.~4(a) the dependence of the precession frequency on the base temperature at low excitation  intensity $I_0$ and on the  laser intensity at low base temperature of 15~K. From the comparison of these two measurements we infer the magnitude of the transient temperature change $\delta T$ as a function of the laser intensity. We note that very similar temperature versus intensity calibration is obtained from the analogous comparison of the intensity dependence of the demagnetizing factor and the temperature dependence of the remanent magnetization measured by SQUID. With the calibration in hand we can proceed to the analysis of the measured easy-axis angles. First, we show in Fig.~4(b) the equilibrium easy-axis orientation  $\varphi_M$ determined from our dynamical measurements at low pump laser intensity and at different base temperatures. We find excellent agreement with the temperature dependent easy-axis angles inferred from SQUID magnetization measurements. These measurements show that with increasing temperature the easy-axis rotates towards the [1$\bar1$0] in-plane diagonal direction. This is because the uniaxial anisotropy component scales with magnetization as $\sim M^2$ while the biaxial component scales as $\sim M^4$ and, therefore, the uniaxial anisotropy gets enhanced relative to the biaxial anisotropy  with increasing temperatures.
\begin{figure}[h!]
\includegraphics[width=1.0\columnwidth,angle=0]{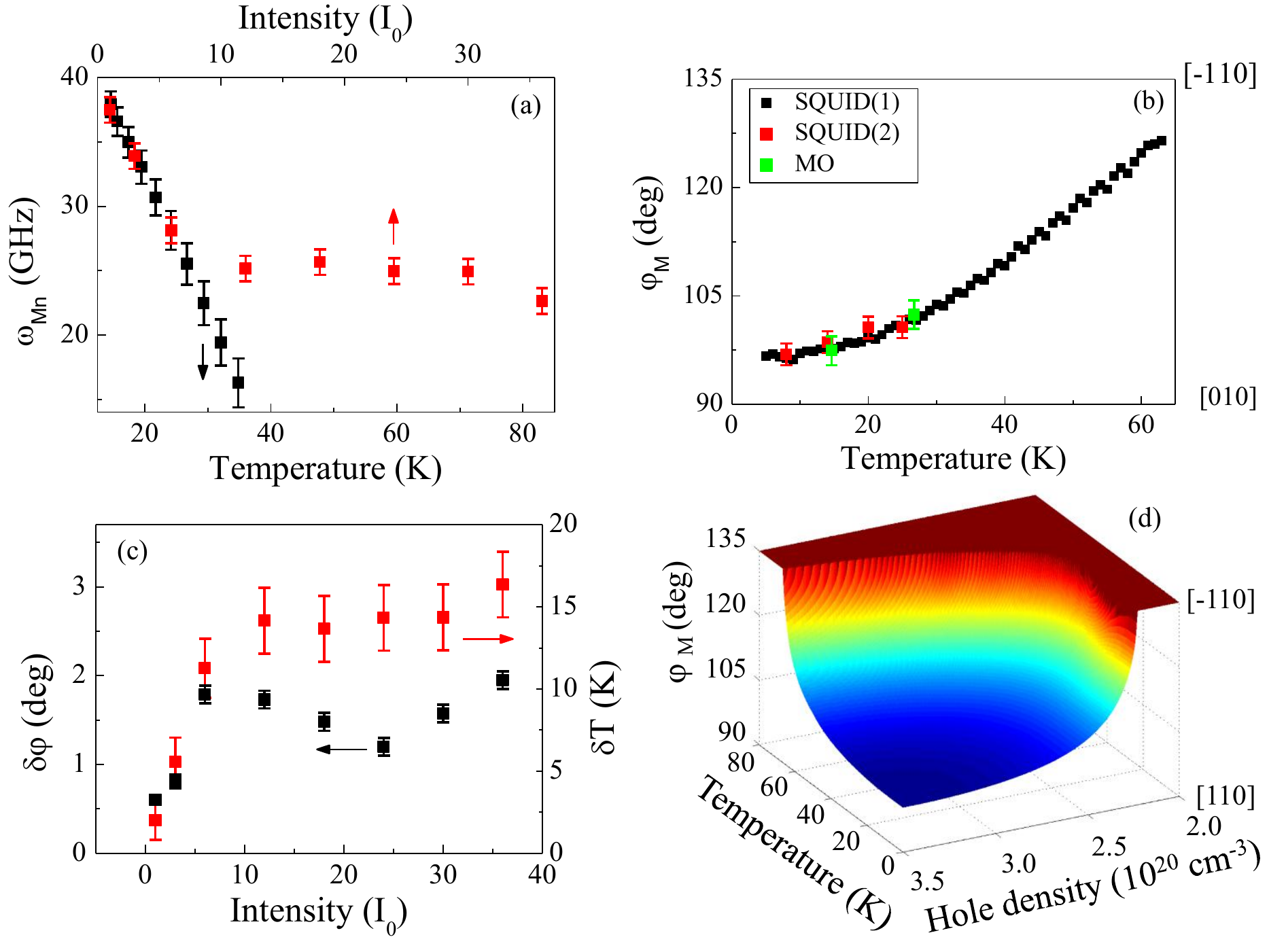}
\caption{{\footnotesize (a) Frequency of  precessing Mn moments in the 3\% doped (Ga,Mn)As measured at $H_{ext}\approx 0$ as a function of base temperature at low excitation intensity $I_0$ and as a function of the laser intensity at low base temperature of 15~K. (b) Temperature dependence of the equilibrium easy-axis orientation $\varphi_M$ determined from remanent magnetization measurements (SQUID(1)), magnetic anisotropy constants inferred from magnetization hysteresis loops (SQUID(2)), and our dynamical magneto-optical measurements at low pump laser intensity $I_0$ (MO). (c) Laser-induced tilt of the easy-axis $\delta\varphi$ compared with the transient temperature increase $\delta T$ (determined from data in (a)) as a function of the pump laser intensity. (d) Microscopic calculations of the temperature and hole density dependent easy-axis angle $\varphi_M$.}}
\label{Fig4}
\end{figure}

In Fig.~4(c) we plot the laser induced tilt of the easy-axis $\delta\varphi$ as a function of the intensity of the pump laser beam. In the same plot we also show the calibration of the transient temperature change $\delta T $ versus intensity inferred from the data in Fig.~4(a). At low laser intensities $\lesssim 7I_0$, $\delta\varphi$ and $\delta T$ are proportional to each other suggesting that thermal effects dominate. This is consistent with the relatively low photo-hole density excited at these low intensities which only reaches $3\times10^{18}$~cm$^{-3}$. Remarkably,  the character of both the temperature  calibration curve and of the $\delta\varphi$ curve changes dramatically at higher intensities. The temperature tends to saturate while $\delta\varphi$ not only varies further with increasing intensity but the sense of the variation reverses, i.e., the easy axis starts to rotate in the opposite direction, back towards the equilibrium angle. Since thermally the easy-axis can only rotate in one direction away from the equilibrium position and since also the transient heating nearly saturates beyond $\approx7I_0$, the origin of $\delta\varphi$ at high laser intensities is non-thermal. 

The photo-generated hole density reaches $10^{19}$~cm$^{-3}$ at high laser intensities for which measurable changes of the easy-axis angle can be readily expected. We therefore attribute the non-thermal effect to the transient increase of the hole density. In Fig.~4(d) we plot microscopic calculations\cite{Zemen:2009_a} of the temperature and hole density dependent easy-axis angle which support this conclusion. The calculations confirm the monotonous rotation of the easy-axis angle towards the in-plane diagonal with increasing temperature and show that an opposite trend can occur in the dependence on the hole concentration. The theoretical results also show that the measured $\delta\varphi\sim-1^{\circ}$ can be explained by the $10^{18}-10^{19}$~cm$^{-3}$ increase of the hole density generated by our high intensity laser pulses.

\vspace*{-.4cm}
\section*{Acknowledgment}
We acknowledge support  from EU ERC Advanced Grant No. 268066, from the Ministry of Education of the Czech Republic Grants No. LC510 and MSM0021620834, from the Grant Agency of the Czech Republic Grant No. 202/09/H041, from the Charles University in Prague Grant No. SVV-2010-261306, and from the Academy of Sciences of the Czech Republic No. AV0Z10100521 and Preamium Academiae.

\newpage

\vspace*{1cm}


\section*{Supplementary material}

# Non-thermal laserinduced precession of magnetization in ferromagnetic semiconductor (Ga,Mn)As: Supplementary material

P. Němec,[1] E. Rozkotová,[1] N. Tesařová,[1] F. Trojánek,[1] K. Olejník,[2]
J. Zemen,[2] V. Novák,[2] M. Cukr,[2] P. Malý,[2] and T. Jungwirth[2,3]

[1] *Charles University in Prague, Faculty of Mathematics and Physics, Ke Karlovu 3, 121 16 Prague 2, Czech Republic*

[2] *Institute of Physics ASCR, v.v.i., Cukrovarnická 10, 16253 Praha 6, Czech Republic*

[3] *School of Physics and Astronomy, University of Nottingham, Nottingham NG72RD, United Kingdom*

## INTRODUCTION

This supplementary material describes a detailed information about the time-resolved magneto-optical experiments performed in a large set of high-quality (Ga,Mn)As epilayers and hybrid structures piezo-stressor/(Ga,Mn)As where *in situ* electrical control of magnetic anisotropy can be achieved. Diluted magnetic semiconductors, with (Ga,Mn)As as the most thoroughly investigated example, are in principle disordered materials. Therefore, a special care has to be taken when generalizing the experimental results obtained in one particular sample to the universal behavior of this material system. Very recently we reported on a systematic study of optical and magneto-optical properties of optimized set of (Ga,Mn)As epilayers spanning the wide range of accessible substitutional $Mn_{Ga}$ dopings [1]. The optimization of the materials in the series, which is performed individually for each nominal doping, minimizes the uncertainties in the experimental sample parameters and produces high quality epilayers which are as close as possible to uniform uncompensated (Ga,Mn)As mixed crystals. For each nominal Mn doping *x*, the growth and post-growth annealing conditions were separately optimized in order to achieve the highest Curie temperature $T_c$ attainable at the particular *x*. The highest $T_c$ criterion was found to lead simultaneously to layers with maximized uniformity and minimized compensation by unintentional impurities and defects [1].

# SAMPLES

The time-resolved magneto-optical experiments described in the main paper were performed in an optimized set of high-quality (Ga,Mn)As epilayers whose selected characteristics are summarized in Tab. I. All samples are in-plane magnets in which the cubic anisotropy competes with an additional uniaxial anisotropy. At very low dopings, the cubic anisotropy dominates and the easy axis align with the main crystal axis [100] or [010]. At intermediate dopings, the uniaxial anisotropy is still weaker but comparable in magnitude to the cubic anisotropy. At very high dopings, the uniaxial anisotropy dominates and the system has one strong easy-axis along the [1-10] in-plane diagonal. We revealed that the laser pulse-induced precession of magnetization can be observed in all (Ga,Mn)As epilayers except in those with very low and very high doping levels where one of the anisotropies strongly dominates. The amplitude of the oscillatory signal depends strongly also on the magnitude and direction of the applied magnetic field.

| sample | $x$ (%) | $d$ (nm) | $T_C$ (K) | $M_s$ (emu/cm$^3$) | Laser-induced precession |
|---|---|---|---|---|---|
| F010 | 1.5 | 20 | 29 | 8.9 | No |
| F008 | 2 | 20 | 47 | 11.6 | Yes |
| F007 | 2.5 | 20 | 60 | 11.5 | Yes |
| F002 | 3 | 20 | 77 | 16.2 | Yes |
| F016 | 3.8 | 20 | 96 | 24.7 | Yes |
| E101 | 4.5 | 19 | 111 | 27.8 | Yes |
| F020 | 5.2 | 20 | 132 | 33.3 | Yes |
| D071 | 7 | 50 | 150 | 47.4 | Yes |
| E115 | 7 | 20 | 159 | 51.0 | Yes |
| E122 | 9 | 20 | 179 | 63.7 | Yes |
| F056 | 14 | 20 | 182 | 78.1 | No |

TABLE I: Table summarizing the basic characteristics of selected samples from the series of optimized materials: $x$ is nominal doping, $d$ is film thickness, $T_C$ is Curie temperature, $M_s$ is saturated magnetic moment. Last column shows if the impact of ultrafast laser pulse induces a precession of magnetization.

We note that even in the two samples with extreme doping levels, where the precession is not observed, the pump-induced changes of the magneto-optical signal (see Fig. 3(b) in the main paper) show an expected doping trend: In the sample with $x = 1.5\%$ the Curie temperature and the background hole concentration are much lower than those in the sample

with $x = 14\%$. Consequently, the same magnitude of the pump-induced transient heating and of the hole concentration change should have a considerably stronger effect in the sample with the lower doping level as was indeed observed.

## PRECESSION OF MAGNETIZATION INDUCED BY LASER PULSES

The impact of a laser pulse on (Ga,Mn)As induces a precession of magnetization by two distinct mechanisms – the helicity-dependent and the helicity-independent. Here we show that they can be directly experimentally separated in a piezo-stressor/(Ga,Mn)As hybrid structure where *in situ* electrical control of magnetic anisotropy can be achieved. This is illustrated in Fig. 1 for a (Ga,Mn)As epilayer with nominal doping $x = 3.8\%$ where the as-measured data for the circularly ($\sigma^+$ and $\sigma^-$) and linearly (*s* and *p*) polarized pump pulses are shown. For the applied voltage U = –150 V, which induces a compressive strain of (-1.5 ± 0.5) × $10^{-4}$ along the piezo-stressor main axis and a tensile strain of (0.3 ± 0.1) × $10^{-4}$ along the minor axis, the precession of magnetization is observed only when circularly-polarized pump pulses are used - see Fig. 1(a). Moreover, a change of the circular polarization helicity results in a phase shift of 180° in the measured signal. On the other hand, for U = +150 V, where a sign of the strains is reversed, pump pulses with any polarization lead to the precession of magnetization but there is not an obvious dependence of the precession phase on the polarization. The co-existence of two rather distinct excitation mechanisms, which both can lead to the precession of magnetization - but with a different initial phase, can be clearly revealed if the polarization-sensitive and polarization-insensitive parts of the signals are computed from the measured data – see Fig. 1(c) – (f). As expected, the polarization-insensitive part of the signal [Fig. 1(e) and (f)] is the same both for linear and circular polarization. This signal, which is connected with a magnetic anizotropy modification due to the pump-induced change of the hole concentration and of the sample temperature, is strongly dependent on the voltage applied to the piezo-stressor. On the contrary, the polarization-sensitive signal contains sizable oscillations only when circular-polarization of pump pulses is used. And remarkably, this signal does not depend significantly on the applied voltage.

The helicity-dependent excitation mechanism is schematically shown in Fig. 2. Before an impact of the pump pulse the magnetization points to the crystallographic direction that corresponds to the minimum energy of the system. If no external magnetic field is applied this direction is determined by a magnetic anisotropy of the sample (easy axis direction). With an external magnetic field $H_{ext}$ applied, this direction is given by an interplay of $H_{ext}$ and the

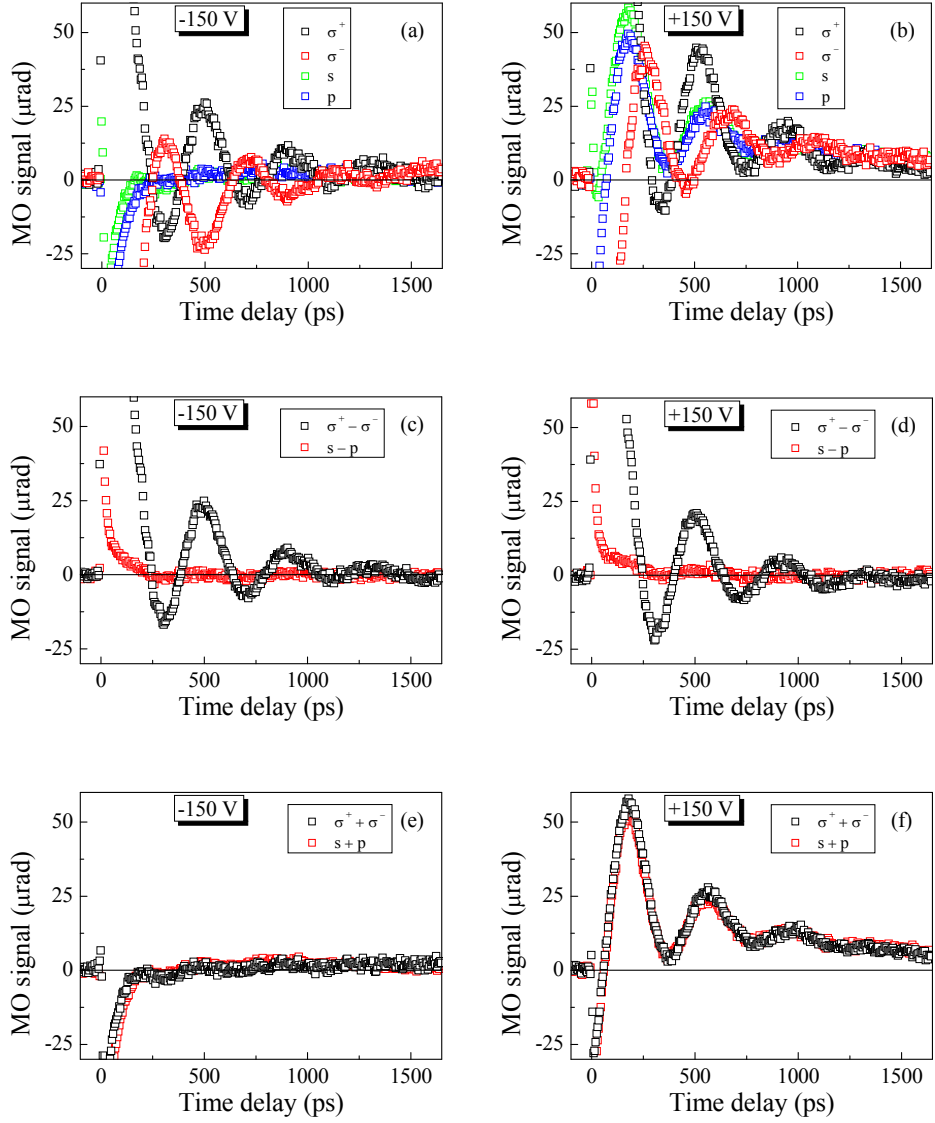

Fig. 1. Laser pulse-induced precession of magnetization in ferromagnetic (Ga,Mn)As epilayer with nominal doping $x = 3.8\%$ (Curie temperature $T_c = 96$ K) attached to a piezo-stressor and measured at $T = 35$ K, $I = 6I_0 =$ 30 μJ.cm$^{-3}$, $\mu_0 H_{ext} = 30$ mT, $\varphi_H = 115°$, $\beta = 25°$. (a) and (b) Data measured for circularly ($\sigma^+$, $\sigma^-$) and linearly ($s$, $p$) polarized pump pulses for applied voltage $U = -150$ V and $U = +150$ V, respectively. Polarization-sensitive [$(\sigma^+ - \sigma^-)/2$ and $(s-p)/2$] and polarization-insensitive [$(\sigma^+ + \sigma^-)/2$ and $(s+p)/2$] parts of these signals are shown in (c), (d) and (e), (f), respectively. (Ga,Mn)As epilayer was mounted to a piezo-stressor in such a way that the angle $\varphi$ between the piezo-stressor main axis and the sample crystallographic direction [100] was 115°; the external magnetic field was applied along the same direction ($\varphi_H = 115°$).

sample anisotropy and can be imagined as a direction of the internal magnetic field $H_{int}$. Due to the conservation of angular momentum, absorption of circularly polarized pump pulse leads to photo-injection of spin-polarized electrons with the spin direction given by the helicity of the circular polarization [4]. The spin-polarized electrons generate an out-of-plane magnetic field $H_{el}$, which can be as high as ~ 20 mT per $10^{18}$ cm$^{-3}$ electron density. $H_{el}$ is added to the

internal field $H_{int}$ and, consequently, it induces a precession of magnetization around its new quasi-equilibrium position. The subsequent recombination and/or spin relaxation of photoinjected electrons lead to the reduction of $H_{el}$ and, finally, to the return of the magnetization equilibrium position to the sample plane. In principle, both the photoinjected electrons and holes can be responsible for this helicity-dependent mechanism. However, the sub-picosecond spin relaxation times of the holes [5] and the fact that the observed characteristic decay time of the photo-electron spin polarization $T_2^*$ coincides with the decay time of the pulse function $\tau_p$ in the Mn moment dynamical signal (see Fig. 1(f) in the main paper) strongly supports the assignment of this affect to the photoinjected electrons.

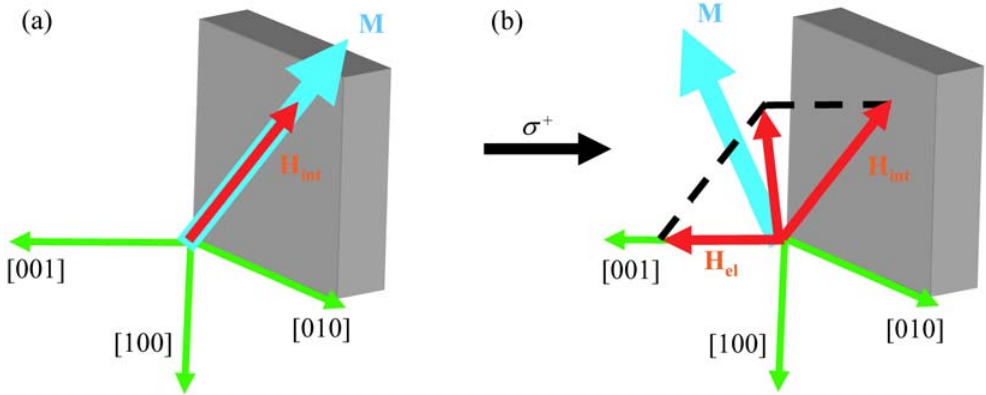

Fig. 2. Schematic illustration of the helicity-dependent mechanism of the precession of magnetization. (a) Before an impact of the pump pulse the magnetization points to the in-plane crystallographic direction that corresponds to the minimum energy of the system (easy axis direction if no external magnetic field is applied, or the position given by the interplay of the sample magnetic anisotropy and the applied external magnetic field - $H_{int}$ that is shown by the red arrow). (b) Absorption of circularly polarized pump pulse photo-generates spin-polarized electrons that produce magnetic field $H_{el}$ which direction is determined by the helicity of the circular polarization. Consequently, the minimum energy of the system is changed, to the position given by a sum of $H_{el}$ and $H_{int}$, that in turn leads to a precession of magnetization around this new quasi-equilibrium position.

The voltage applied in the studied piezo-stressor/(Ga,Mn)As hybrid structure significantly influences the strain in the sample. This leads to a suppression of the helicity-independent mechanism by tuning the sample magnetic anisotropy to the state where it does not depend on the pump-induced change of the hole concentration and the sample temperature (which happens for U = −150 V in this particular sample). Moreover, the strain can also modify the precession frequency as shown in Fig. 3. The frequency of the magnetization precession is given by the magnetic anisotropy of (Ga,Mn)As as well as by the magnitude and direction of the external magnetic field [6]. If no magnetic field is applied, the change of the magnetic anisotropy due to the piezo-controlled strain is clearly apparent in the precession

frequency – see Fig. 3 for fields up to 10 mT. On the contrary, above 10 mT, when the frequency is given mainly by the applied magnetic field, the voltage applied to the piezo does not have any sizable effect.

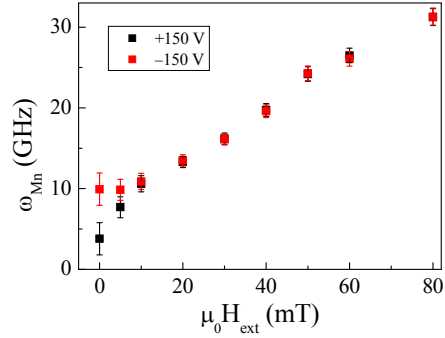

Fig. 3. Influence of the external magnetic field $H_{ext}$ on the precession angular frequency $\omega_{Mn}$ for two voltages applied to the piezo-stressor/(Ga,Mn)As hybrid structure; $I = 6I_0 = 30$ μJ.cm$^{-3}$, $\mu_0 H_{ext} = 30$ mT, $\varphi_H = 115°$, $\beta = 25°$.

## DYNAMICS OF MAGNETIZATION VS. DYNAMICS OF MAGNETO-OPTICAL SIGNAL

Magneto-optics deals with phenomena induced by interaction between light and a matter exposed to a magnetic field (external or internal). Reflection of linearly polarized light from the magnetic medium (with a magnetization along the light propagation) leads to the rotation of light polarization $\phi_1$ and to the change of its ellipticity $\phi_2$. For the polar Kerr effect they can be expressed in the following form [7]:

$$\phi_1 = f_{\phi 1} \cdot M , \tag{1a}$$

$$\phi_2 = f_{\phi 2} \cdot M , \tag{1b}$$

where $M$ is the magnetization, $f_{\phi 1}$ and $f_{\phi 2}$ are functions that depend on the electronic properties of the material and that can be expressed in terms of the refractive index and the absorption coefficient. Correspondingly, the light-induced dynamical change of $\phi_1$ and $\phi_2$ consists of two components

$$\Delta\phi_1(t) \approx f_{\phi 1} \cdot \Delta M(t) + \Delta f_{\phi 1}(t) \cdot M , \tag{2a}$$

$$\Delta\phi_2(t) \approx f_{\phi 2} \cdot \Delta M(t) + \Delta f_{\phi 2}(t) \cdot M , \tag{2b}$$

6

where only the first term (the so-called "magnetic part" of the signal) reflects the magnetization dynamics while the second term (the so-called "optical part" of the signal) is a consequence of the pump pulse-induced change of the complex index of refraction of the sample. Consequently, the dynamics of both the rotation and ellipticity has to be measured and compared before the obtained magneto-optical signal is attributed to the magnetization dynamics. In the case of (Ga,Mn)As, the situation is further complicated by the fact that not only the polar Kerr effect but also the magnetic linear dichroism contribute to the measured magneto-optical signal (see below). Nevertheless, the above discussion remains qualitatively valid also in this case. The laser pulse-induced change of the rotation and ellipticity measured in a (Ga,Mn)As epilayer with nominal doping $x = 3\%$ (Curie temperature $T_c = 77$ K) is shown in Fig. 4. The curves are very similar except for short time delays. This implies that the "optical part" of the signal has a sizable contribution only in a time range up to ≈ 30 ps where the reflectivity (i.e., the complex index of refraction) is modified considerably by the pump pulse (see the inset of Fig 4).

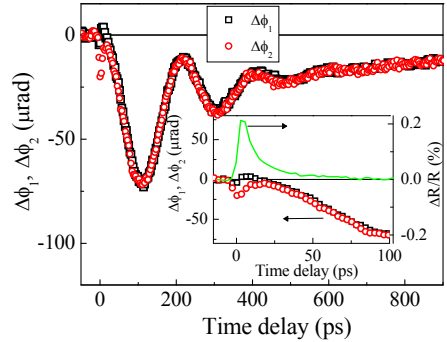

Fig. 4. Comparison of the laser pulse-induced change of the rotation $\Delta\phi_1$ and ellipticity $\Delta\phi_2$ measured in a (Ga,Mn)As epilayer with nominal doping $x = 3\%$ (Curie temperature $T_c = 77$ K); $T = 15$ K, $I = 6I_0 = 30$ μJ.cm$^{-3}$, $\mu_0 H_{ext} \approx 0$ mT, $\beta = 105°$. Inset: Detail of the measured dynamics of $\Delta\phi_1$ and $\Delta\phi_2$ (points) together with the pulse-induced reflectivity change $\Delta R/R$ (line).

The measured dynamics of the reflectivity change $\Delta R/R$ provides us information also about the characteristic relaxation and recombination times of electrons in the investigated samples [8]. From the measured data we can conclude that the population of photo-generated free electrons decays within ≈ 30 ps after the photo-injection. This rather short lifetime of free electrons is similar to that reported for the low temperature grown GaAs (LT-GaAs), which is generally interpreted as a result of a high concentration of nonradiative recombination centers induced by the low temperature growth mode of the MBE [9]. The nonradiative recombination of photo-injected electrons is accompanied by an emission of phonons (i.e., by

7

a heating of the sample). Consequently, the measured decay time of $\Delta R/R$ corresponds to a rise time of the laser-induced transient change of the sample temperature $\delta T$.

## STATIC MAGNETO-OPTICAL SIGNALS IN (GA,MN)AS

In (Ga,Mn)As there are two magneto-optical (MO) effects that are responsible for the measured signal. In the following we will concentrate on the rotation of the polarization plane of the reflected linearly polarized light but the same apply also for the change of the light ellipticity. We will also limit the discussion to the case when the incident light is close to the normal incidence (in our experiment the angle of incidence is 2° and 8° for pump and probe pulses, respectively). The first of the effects is the well-known polar Kerr effect (PKE), which is sometimes called magneto-optical Kerr effect (MOKE), where the rotation of polarization occurs due to the different index of refraction for $\sigma^+$ and $\sigma^-$ circularly polarized light propagating *parallel to the direction of magnetization* - see Fig. 5(a). Consequently, the rotation of light polarization $\Delta\beta$ is proportional to the projection of magnetization to the direction of light propagation (see Fig. 1(a) in the main paper for the definition of the coordinate system)

$$MO^{PKE} \equiv \Delta\beta^{PKE} \equiv \beta' - \beta = P^{PKE} \frac{M_z}{M_s} = P^{PKE} \cos\theta_M, \tag{3}$$

where $\beta$ and $\beta'$ describes the orientation of the input and output linear polarization [see Fig. 6(a)], $P^{PKE}$ is the corresponding magneto-optical coefficient of the sample, $M_s$ and $M_z$ are the magnitude and $z$ component of magnetization, and $\theta_M$ describes the out-of-plane orientation of magnetization, respectively. Here we adopted the following sign convention: If light is reflected along the direction of magnetization, the value $P^{PKE} > 0$ corresponds to a counterclockwise rotation of incident polarization (i.e., $\Delta\beta > 0$) when viewed by an observer facing the sample – see Fig. 6(a). We note that this MO effect is linear in magnetization (i.e., the sign of $\Delta\beta$ is changed when the direction of magnetization is reversed) and that the value of $\Delta\beta^{PKE}$ does not dependent on $\beta$.

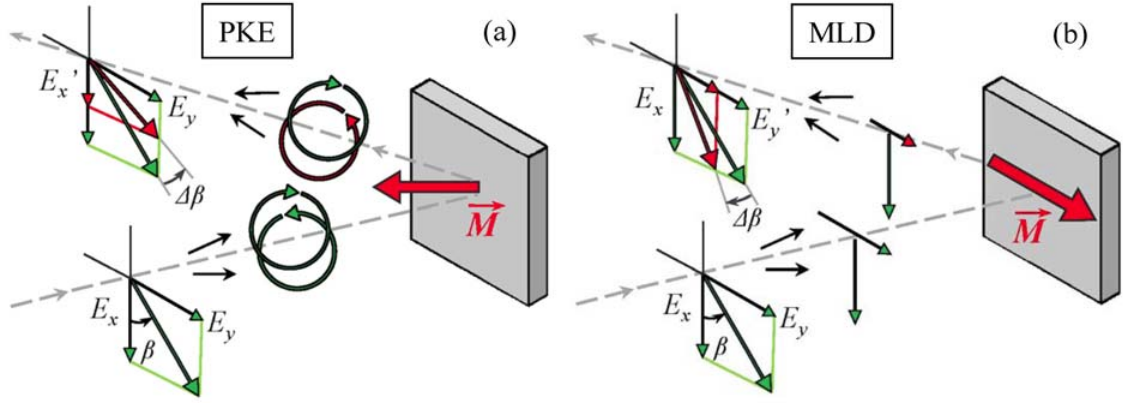

Fig. 5. Schematic illustration of the two magneto-optical effects in (Ga,Mn)As that are responsible for a rotation of the polarization plane $\Delta\beta$ of reflected light at normal incidence. (a) Polar Kerr effect (PKE) that is due to the different index of refraction for $\sigma^+$ and $\sigma^-$ circularly polarized light propagating parallel to the direction of magnetization $M$. (b) Magnetic linear dichroism (MLD) that is due to the different absorption (reflection) coefficient for light linearly polarized parallel and perpendicular to $M$ if the light propagates perpendicular to the direction of $M$.

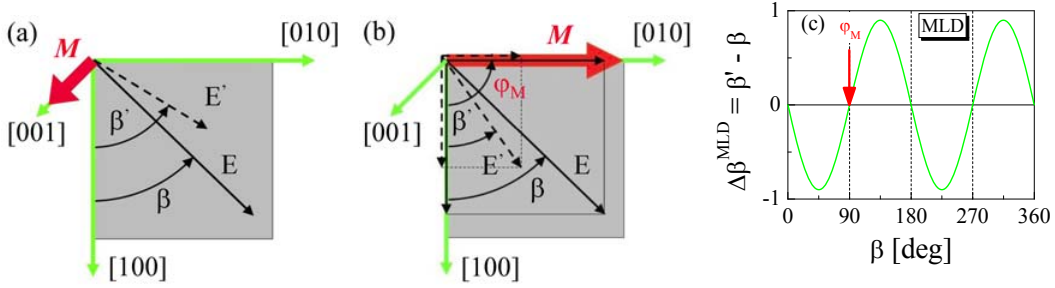

Fig. 6. Polarization dependence of magneto-optical effects. (a) PKE is proportional to the out-of-plane projection of magnetization; for $P^{PKE} > 0$ and $M_z > 0$ this MO effect leads to $\Delta\beta > 0$ for any $\beta$. (b) MLD is sensitive to the in-plane projection of magnetization; the magnitude and sign of $\Delta\beta$ is a harmonic function of $\beta$ as described by Eq. (7) and schematically illustrated in (c) where the vertical red arrow depicts the assumed position of the magnetization.

The second MO effects is the magnetic linear dichroism (MLD) [10], which originates from to the different absorption (reflection) coefficient for light linearly polarized parallel and perpendicular to $M$, that occurs if the light propagates *perpendicular to the direction of magnetization M* - see Fig. 5(b). To derive the rotation of light polarization due to MLD we first suppose that the magnetization is located in the sample plane with a position characterized by an azimuthal angle $\varphi_M$. We can express the projections of the incident electric field amplitude $E$ parallel to magnetization ($E^{\parallel}$) and perpendicular to magnetization ($E^{\perp}$) using $\varphi_M$ and the input polarization orientation $\beta$

$$E^{\parallel} = E\cos(\varphi_M - \beta), \tag{4a}$$

$$E^{\perp} = E\sin(\varphi_M - \beta). \tag{4b}$$

9

The same can be done for the reflected electric field amplitude $E'$. If we now consider that $E^{\parallel}$ ($E^{\perp}$) is reflected from (Ga,Mn)As with the amplitude reflection coefficient $a$ ($b$), we obtain

$$tg(\varphi_M - \beta) = \frac{E^{\perp}}{E^{\parallel}}, \tag{5a}$$

$$tg(\varphi_M - \beta') = \frac{bE^{\perp}}{aE^{\parallel}}, \tag{5b}$$

from which the rotation of light polarization $\Delta\beta \equiv \beta' - \beta$ can be easily derived

$$tg\Delta\beta = \frac{(a-b)tg(\varphi_M - \beta)}{a + b[tg(\varphi_M - \beta)]^2}. \tag{6}$$

If we now assume that $a/b \approx 1$ (i.e., that $\Delta\beta$ is small) we obtain

$$\Delta\beta = P^{MLD} \sin 2(\varphi_M - \beta), \tag{7}$$

where the magneto-optical coefficient $P^{MLD}$ is defined as

$$P^{MLD} = 0.5\left(\frac{a}{b} - 1\right). \tag{8}$$

In a more general case, when magnetization has an arbitrary orientation, which is characterized by $\varphi_M$ and $\theta_M$, the rotation of light polarization by MLD is given by

$$MO^{MLD} \equiv \Delta\beta^{MLD} \equiv \beta' - \beta = P^{MLD} \sin\theta_M \sin 2(\varphi_M - \beta). \tag{9}$$

The total MO response of any (Ga,Mn)As sample is given by a sum of contributions due to PKE and MLD:

$$MO^{stat} = MO^{PKE} + MO^{MLD} = P^{PKE} \cos\theta_M + P^{MLD} \sin\theta_M \sin 2(\varphi_M - \beta) \tag{10}$$

The magnitude of $P^{PKE}$ and $P^{MLD}$ can be directly measured if the magnetization is oriented by a strong external magnetic field to the out-of-plane ($\theta_M = 0°$) and in-plane ($\theta_M = 90°$) positions, respectively. As an example, we show in Fig. 7(a) the spectral dependence of $P^{PKE}$ and $P^{MLD}$ measured in a (Ga,Mn)As epilayer with nominal doping $x = 7\%$ (Curie temperature $T_c = 150$ K). We note that even though the magnitude of $P^{PKE}$ is typically larger than that of $P^{MLD}$, there exists a relatively broad spectral region around 1.6 eV where they are comparable. In Fig. 7(b) the measured temperature dependence of $P^{MLD}$ is compared with $M^2$, which was measured by SQUID, that confirms the expected [10] quadratic dependence of $P^{MLD}$ on $M$.

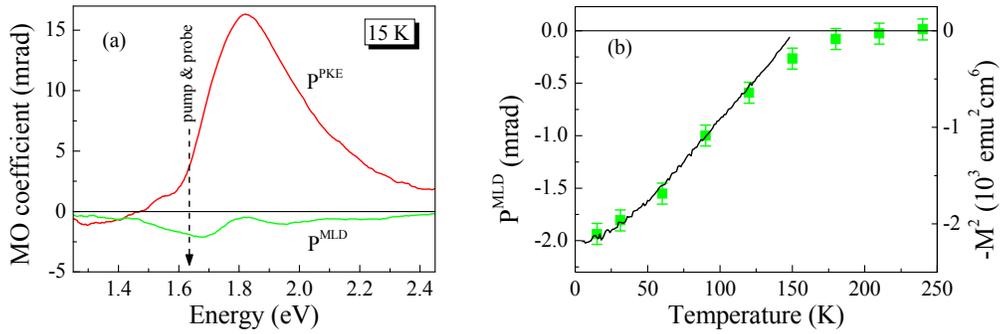

Fig. 7. (a) Spectral dependence of magneto-optical coefficients $P^{PKE}$ and $P^{MLD}$ measured in a (Ga,Mn)As epilayer with nominal doping $x = 7\%$ (Curie temperature $T_c = 150$ K) at temperature 15 K; the vertical arrow shows schematically the spectral position of laser pulses used for the pump & probe experiment. (b) Temperature dependence of $P^{MLD}$ measured at 1.64 eV (points) and temperature dependence of $(-1) \times M^2$ (line).

## DYNAMIC MAGNETO-OPTICAL SIGNALS IN (GA,MN)AS

In the previous chapter we have shown that in a general case both PKE, which is sensitive to the out-of-plane component of magnetization, and MLD, which is sensitive to the in-plane component of magnetization, contribute to the measured magneto-optical signals in (Ga,Mn)As. All the investigated (Ga,Mn)As samples with nominal doping ranging from 1.5% to 14% are in-plane magnets (i.e., $\theta_M = 90°$). Consequently, in the equilibrium conditions the static MO signal is only due to MLD:

$$MO^{stat} = P^{MLD} \sin 2(\varphi_M - \beta) \tag{11}$$

The impact of a strong pump laser pulse modifies the properties of the sample that finally leads to the precession of magnetization around the new quasi-equilibrium position.

The measured pump-induced dynamical change of the magneto-optical signal, $\delta MO$, can be fitted well by the phenomenological equation,

$$\delta MO(t) = A\cos(\omega_{Mn}t + \Delta)e^{-t/\tau_G} + Ce^{-t/\tau_p} + D\cos(\omega_L t)e^{-t/T_2^*}, \qquad (12)$$

where $A$ and $C$ are the amplitudes of the oscillatory and pulse function, respectively, $\omega_{Mn}$ is the ferromagnetic moment precession frequency, $\Delta$ is the phase factor, $\tau_G$ is the Gilbert damping time, and $\tau_p$ is the pulse function decay time. The last term in Eq. (12) describes the signal due to the photo-exited spin-polarized conduction band electrons, which is present only in the helicity-dependent signal - see the main paper, and, therefore, we will not consider it in the following analysis of the MO signal that is connected with the ferromagnetically coupled *Mn* spins. The pulse function in $\delta MO$ signal is a transient non-oscillatory change of the static signal $MO^{stat}$. In fact, there are two distinct contributions to this signal. Firstly, there is a contribution due to a change of the magnetization position (the "shift" signal in the following), which corresponds to a derivative of Eq. (10) with respect to a small change of $\varphi_M$ and $\theta_M$. Secondly, the pump-induced demagnetization of the material [11] reduces also the static MO response (the "demagnetization" signal), which is in the investigated samples with the in-plane anisotropy (i.e., $\theta_M = 90°$) given by Eq. (11). If we assume, for simplicity, that both these signals have the same dynamics, which is presumably dominated by a dissipation of heat from the irradiated spot on the sample with a locally increased temperature $\delta T$, we have the following equation for the measured amplitude of the pulse function $C$:

$$C = C^{shift} + C^{demag} = -\delta\theta\, P^{PKE} + \delta\varphi\, P^{MLD} 2\cos 2(\varphi_M - \beta) - \delta p\, P^{MLD}\sin 2(\varphi_M - \beta), \qquad (13)$$

where $\delta\theta$ and $\delta\varphi$ describe the out-of-plane and in-plane movement of the quasi-equilibrium position of magnetization, respectively, and the demagnetization factor $\delta p$ characterizes a reduction of the magnetization magnitude. In fact, Eq. (13) is of fundamental importance for the analysis of the measured MO signals because it enables to determine experimentally if the precession of magnetization is triggered by the out-of-plane ($\delta\theta$) or by the in-plane ($\delta\varphi$) movement of the quasi-equilibrium position of magnetization along which the magnetization precesses. If the out-of-plane movement dominates, $C$ does not depend on $\beta$. On the other hand, if the in-plane movement dominates, $C$ is a harmonic function of $\beta$ [see Eq. (13)]. If

12

both movements are comparable, $C$ is again a harmonic function of $\beta$ but in this case there is an offset. This is shown in Fig. 2(d) in the main paper - the quasi-equilibrium position of magnetization is tilted in the out-of-plane direction for the helicity-dependent signal and in the in-plane direction for the helicity-independent signal.

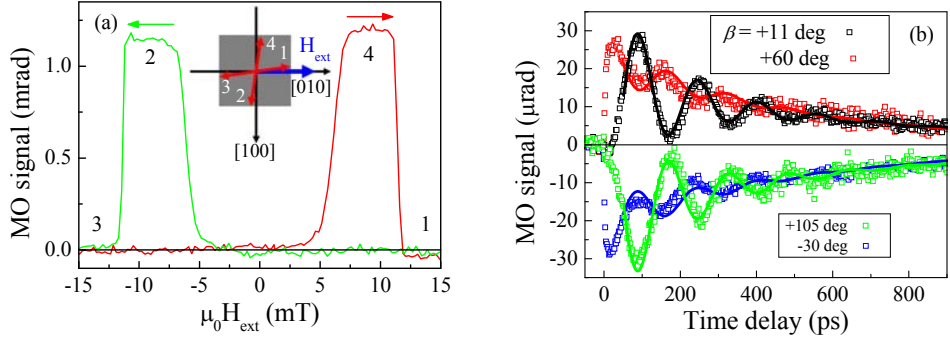

Fig. 8. Magneto-optical signal measured at temperature 15 K in a (Ga,Mn)As epilayer with nominal doping $x$ = 3% (Curie temperature $T_c$ = 77 K). (a) $M$-shaped hysteresis loop that is a signature of the existence of four energetically equivalent easy axis positions of magnetization [10], which are schematically labeled "1" to "4" in the inset. (b) Dynamics of the helicity-independent MO signal induced by an impact of pump pulse on the sample with a magnetization in the position "1" that was measured by probe pulses with different $\beta$ (points). Lines are fits by Eq. (12).

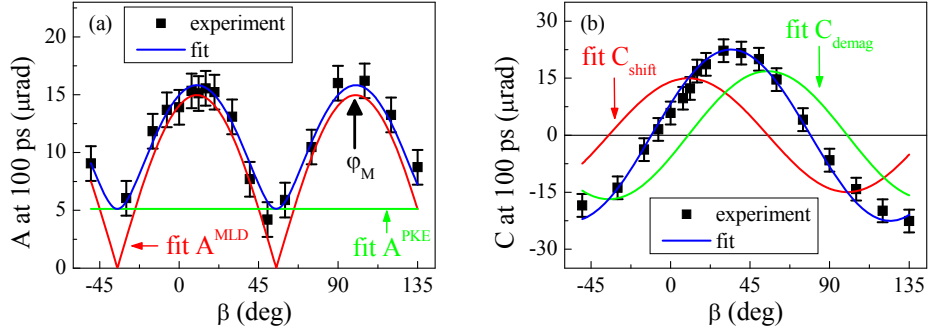

Fig. 9. Polarization dependence of the amplitudes of the oscillatory part $A$ (a) and of the pulse function $C$ (b) that was obtained by fitting the dynamics shown in Fig. 8(b) by Eq. (12) with a fixed values $\omega_{Mn}$ = 33.9 GHz, $\tau_G$ = 165 ps, $\tau_p$ = 880 ps, and $D$ = 0 µrad; we show the values of $A$ and $C$ at time delay of 100 ps to account for the difference in $\tau_G$ and $\tau_p$ (points). Lines are results of simultaneous fits of $A(\beta)$ by Eq. (16) and $C(\beta)$ by Eq. (13). The vertical arrow in (a) depicts the measured easy axis position in the sample without the pump pulse. These figures are re-plotted from Fig. 3(c) and Fig. 3(d) in the main paper.

Also the oscillatory MO signal contains a signal due to the out-of-plane motion of the magnetization, which is sensed by PKE, and a signal due to the in-plane movement of magnetization, which is sensed by MLD. Due to the precessional motion of magnetization these signals are phase shifted for 90° and the total amplitude of the oscillatory MO signal $A$ is given by

13

$$A(\beta) = \sqrt{\left[A^{MLD}(\beta)\right]^2 + \left[A^{PKE}\right]^2}. \tag{14}$$

This equation explains why $A$ depends strongly on the orientation of the probe polarization $\beta$ (see Fig. 2 and 3 in the main paper). The MO signal due to the out-of-plane projection of magnetization (with an amplitude $A^{PKE}$) does not depend on $\beta$ but the MO signal due to the in-plane projection (with an amplitude $A^{MLD}$) is very sensitive to $\beta$. In Fig. 8 we show, as an example, the data measured in a (Ga,Mn)As epilayer with nominal doping $x = 3\%$ (Curie temperature $T_c = 77$ K). Prior to the time-resolved experiment, we prepared the magnetization in a state close to [010] crystallographic direction [easy axis position labeled "1" in Fig. 8(a)]. An impact of the pump laser pulse changes the easy axis position and, consequently, triggers the precession of magnetization around this new quasi-equilibrium position. The resulting precessional MO signal strongly depends on the orientation of linear polarization of probe pulses $\beta$ as shown in Fig. 8(b) where the helicity-independent part of the MO signal is depicted. All the measured dynamics can be fitted well by Eq. (12) with a one set of parameters $\omega_{Mn}$, $\tau_G$ and $\tau_p$. The dependences $A(\beta)$ and $C(\beta)$ obtained by this fitting procedure are displayed in Fig. 9(a) and (b). From these graphs two conclusions can be immediately obtained. Firstly, the dependence $C(\beta)$ is a harmonic function of $\beta$ with no significant offset. This show, as discussed above, that the precession is started by the in-plane movement of the easy axis. Secondly, the position of the maximum in the dependence $A(\beta)$ at $\beta \approx 100°$ corresponds to the equilibrium position $\varphi_M$ of the easy axis in the sample (i.e., its position without the pump pulse). The second conclusion immediately follows from the fact, that the $\beta$ dependence of $A$ comes from the MO signal induced by a change of the in-plane projection of magnetization, which is detected by MLD. And from Fig. 6(c) it is clearly apparent that the strongest change of the MO signal due to an in-plane movement of magnetization is observed when probe pulses are polarized along the magnetization or perpendicular to it (i.e., when the derivative of Eq. (7) with respect to $\varphi_M$ is the largest). We recall that prior to this measurement we prepared the magnetization in a state close to [010] crystallographic direction.

The laser pulse-induced shift of the easy axis position is usually much faster than the precessional period $T_{osc} = 2\pi/\omega_{Mn}$ and the Gilbert damping time $\tau_G$. For example, in the data shown in Fig. 8(b) we have $T_{osc} = 158$ ps and $\tau_G = 165$ ps that is considerably longer than the rise time of the laser-induced transient change of the sample temperature $\Delta T$, which is $\approx 30$ ps

(see Fig. 4), and the hole concentration $\Delta p$, which is expected to be quasi-instantaneous. Under these conditions, the initial amplitude of the oscillations should be equal to the in-plane movement of the easy axis:

$$A^{MLD} = C^{shift} = \delta\varphi P^{MLD} 2\cos 2(\varphi_M - \beta). \tag{15}$$

Substituting Eq. (15) to Eq. (14) yields

$$A(\beta) = \sqrt{\left[\delta\varphi P^{MLD} 2\cos 2(\varphi_M - \beta)\right]^2 + \left[A^{PKE}\right]^2}. \tag{16}$$

Consequently, Eq. (16) and (13) can be used to fit the measured dependences $A(\beta)$ and $C(\beta)$, respectively (solid lines in Fig. 9). As an input to the fitting procedure we used the *independently measured* value of the MO constant $P^{MLD} = -0.9 \pm 0.1$ mrad and we obtained $\varphi_M = 100 \pm 2°$, $\delta\varphi = +0.45 \pm 0.05°$, $\delta\theta = 0°$, and $\delta p = 1.9 \pm 0.3\%$. Here we would like to stress that by the simultaneous fitting of $A(\beta)$ and $C(\beta)$ we can determine not only the magnitude of the easy axis tilt in the sample plane, $\delta\varphi$, but also its sign.

## NUMERICAL MODELLING OF DYNAMIC MAGNETO-OPTICAL SIGNAL BY LLG EQUATION

To corroborate the model presented above, we performed also an extensive numerical modeling the measured precessional MO signal by the Landau-Lifshitz-Gilbert (LLG) equation. We used LLG equation in spherical coordinates where the time evolution of magnetization magnitude $M_s$ and orientation, which is characterized by the polar $\theta$ and azimuthal $\varphi$ angles, is given by:

$$\frac{dM_s}{dt} = 0, \tag{17}$$

$$\frac{d\theta}{dt} = -\frac{\gamma}{(1+\alpha^2)M_s}\left(\alpha \cdot A + \frac{B}{\sin\theta}\right), \tag{18}$$

$$\frac{d\varphi}{dt} = \frac{\gamma}{(1+\alpha^2)M_s \sin\theta}\left(A - \frac{\alpha \cdot B}{\sin\theta}\right), \tag{19}$$

where α is the Gilbert damping parameter and $\gamma$ is the gyromagnetic ratio. Functions $A = dF/d\theta$ and $B = dF/d\varphi$ are the derivatives of the energy density functional $F$ with respect to $\theta$ and $\varphi$, respectively. We expressed $F$ in a form:

$$F = M\left[\begin{array}{l} K_c \sin^2\theta\left(\frac{1}{4}\sin^2 2\varphi \sin^2\theta + \cos^2\theta\right) - K_{[001]}\cos^2\theta - \frac{K_{[110]}}{2}\sin^2\theta(1-\sin 2\varphi) - \\ -H_{ext}[\cos\theta\cos\theta_H + \sin\theta\sin\theta_H \cos(\varphi - \varphi_H)] \end{array}\right], \tag{20}$$

where $K_c$, $K_{[001]}$ and $K_{[110]}$ are constants that characterize the magnetic anisotropy in (Ga,Mn)As [12] and $H_{ext}$ is the external magnetic field whose orientation is given by the angles $\theta_H$ and $\varphi_H$. We would like to stress that our approach is considerably distinct from that used up to now for a modeling of the laser pulse-induced precession of magnetization in (Ga,Mn)As [13, 14] where the magnetic anisotropy of the material was characterized by *one effective magnetic field, which was a fitting parameter*. Instead, we used a realistic model of the magnetic anisotropy in (Ga,Mn)As [12] and we *measured independently* the corresponding anisotropy constants in the sample by SQUID.

For the numerical modeling of laser pulse-induced precession of magnetization we used the scenario that is schematically depicted in Fig. 10. Before an impact of the pump pulse the magnetization points to the easy axis direction. Absorption of the laser pulse leads to the photo-injection of electron-hole pairs with a concentration $\delta n = \delta p$. The subsequent fast nonradiative recombination of photo-injected electrons induces a transient increase of the lattice temperature $\delta T$ (within tens of picoseconds after the impact of the pump pulse). The laser-induced change of the lattice temperature and of the hole concentration lead to a change of the easy axis position. Consequently, magnetization starts to follow the easy axis shift by the precessional motion. Finally, dissipation of the heat and recombination of the excess holes lead to the return of the easy axis to the equilibrium position and the precession of magnetization is stopped by the Gilbert damping.

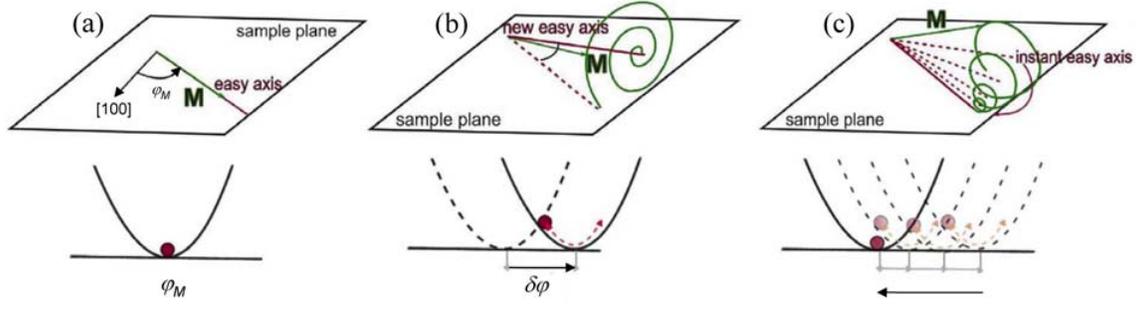

Fig. 10. Schematic illustration of the helicity-independent laser pulse-induced precession of magnetization. (a) In the equilibrium the magnetization points to the easy axis direction, which is located in the sample plane at azimuthal angle $\varphi_M$. (b) Impact of a pump pulse induces a transient increase of the lattice temperature $\Delta T$ and of the hole concentration $\Delta p$ that lead to a change of the easy axis position and, consequently, to the precession of magnetization. (c) Dissipation of the heat and recombination of the excess photoinjected holes lead to the return of the easy axis to the equilibrium position. Simultaneously with this, the precession of magnetization is stopped by the Gilbert damping.

To model the measured data, we first computed from LLG equation the time-dependent deviations of the spherical angles [$\delta\theta(t)$ and $\delta\varphi(t)$] from the corresponding equilibrium values ($\theta_M = 90°$, $\varphi_M \approx 100°$). Then we calculated how such changes of $\theta$ and $\varphi$ modify the magneto-optical response of the sample, which is the signal that we detect experimentally:

$$\delta MO(t) = \frac{\partial(MO^{stat})}{\partial \theta}\delta\theta(t) + \frac{\partial(MO^{stat})}{\partial \varphi}\delta\varphi(t) + \delta MO^{demag}(t) \ . \qquad (21)$$

The first two terms in Eq. (21) are connected with the movement of magnetization; the static magneto-optical response of the sample ($MO^{stat}$) is given by Eq. (10). The last term in Eq. (21) describes a change of the static magneto-optical response of the sample due to the laser-pulses induced demagnetization, as was already discussed in the previous chapter. In Fig. 11(a) we show the precession data measured by probe pulses with different polarization angle $\beta$ in a (Ga,Mn)As epilayer with nominal doping $x = 3\%$ (points) – we note that this is the same data set that was already shown in Fig. 8. The lines in Fig. 11(a) are fits by our numerical model and the corresponding dynamics of $\delta\theta(t)$ and $\delta\varphi(t)$, which are the same for all the curves in Fig. 11(a), are shown in Fig. 11(b). And we again stress that the magnetic anisotropy constants ($\mu_0 K_c = 68.8$ mT, $\mu_0 K_{[110]} = 20.4$ mT and $\mu_0 K_{[001]} = -115$ mT) and the magneto-optical constants ($P^{PKE} = -0.24$ mrad, $P^{MLD} = -0.9$ mrad) of the sample were not fitted – they were measured in independent experiments. Finally, we would like to mention

that the computed magnitude of the in-plane movement of magnetization is in a perfect agreement with the value obtained from the model based on $A(\beta)$ and $C(\beta)$ dependences, which was discussed in the previous chapter.

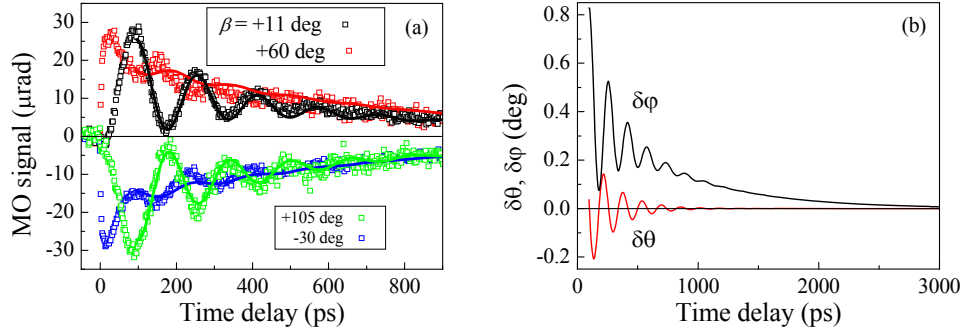

Fig. 11. Modeling of the magneto-optical signal measured at temperature 15 K in a (Ga,Mn)As epilayer with nominal doping $x = 3\%$ (Curie temperature $T_c = 77$ K) using LLG equation. (a) Dynamics of the helicity-independent MO signal measured by probe pulses with different $\beta$ (points). Lines are fits by the numerical model described in the text. (b) Time evolution of $\delta\theta(t)$ and $\delta\varphi(t)$ that were used to model the data shown in (a).



\end{document}